\documentclass{article}

\usepackage[final, nonatbib]{nips_2016}

\usepackage[utf8]{inputenc}		
\usepackage[T1]{fontenc}    	
\usepackage{url}            	
\usepackage{booktabs}       	
\usepackage{amsfonts}       	
\usepackage{amsmath}			
\usepackage{nicefrac}       	
\usepackage{microtype}      	
\usepackage{csquotes}			
\usepackage{graphicx}
\usepackage{comment}
\usepackage{sidecap}			
\title{Real-time eSports Match Result Prediction}

\author{
  Yifan Yang \qquad Tian Qin \qquad Yu-Heng Lei  \\
  Language Technologies Institute\\
  Carnegie Mellon University\\
  \texttt{\{yifany1, tq, yuhengl1\}@andrew.cmu.edu} \\
}

\begin{document}
\maketitle

\begin{abstract}
In this paper, we try to predict the winning team of a match in the multiplayer eSports game Dota 2. To address the weaknesses of previous work, we consider more aspects of prior (pre-match) features from individual players' match history, as well as real-time (during-match) features at each minute as the match progresses. We use logistic regression, the proposed Attribute Sequence Model, and their combinations as the prediction models. In a dataset of 78362 matches where 20631 matches contain replay data, our experiments show that adding more aspects of prior features improves accuracy from 58.69\% to 71.49\%, and introducing real-time features achieves up to 93.73\% accuracy when predicting at the 40th minute.
\end{abstract}

\section{Introduction}
\label{chap:intro}
Predicting match results in sports games has always been a popular research topic in machine learning. Indeed, sports analytics has been an emerging field in many professional sports games to help decision making\footnote{http://www.forbes.com/sites/leighsteinberg/2015/08/18/changing-the-game-the-rise-of-sports-analytics/}. In recent years, electronic sports (a.k.a. eSports), a form of competitions on multiplayer video games, have also been recognized as legitimate sports. For example, Dota 2, a multiplayer online battle arena game, is one of the most active games in the eSports industry, with over one million concurrent players on the Steam gaming platform. The prize of professional Dota 2 tournaments has already passed \$65 million by June 2016.\footnote{https://techvibes.com/2016/06/21/how-videogames-became-a-sport-and-why-theyre-here-to-stay-hint-money} Just as sports analytics has been used in decision making of professional sports, it is foreseeable that “eSports analytics” will also be useful for players in professional eSports competitions, live streaming media that cover these competitions, or eSports game developers.

In this paper, we try to predict the winning team of a match in the multiplayer eSports game Dota 2, in which a match consists of two teams named “Radiant” and “Dire”, and each team consists of five players. Before a match begins, each player selects a unique “hero” character to be controlled in the match. As the match progresses, the hero can farm gold or obtain experience to level up via combats against rival heroes. The winning criteria for a team is to destroy the opponent team.

Our goal to predict the winning team of Dota 2 has two stages. In our first stage, we predict the result before a match begins. Because previous work \cite{Stanford2015, UCSD2015} used very limited aspects of features, we extract more aspects of features from the available game data. In addition, previous work only consider pre-match information, while the most useful information is typically generated during the game. This motivates our second stage to further introduce real-time gameplay data to predict during a match. Therefore, the contributions of this paper are:
\begin{enumerate}
\item Consider more aspects of prior features (\textit{before a match begins}) from individual players' match history instead of only hero features to improve prediction accuracy.
\item Consider real-time gameplay features (\textit{during a match}) to predict the winning team as the match progresses (e.g. predicting at each 1-minute interval).
\end{enumerate}

\begin{SCfigure}
\centering
\includegraphics[width=0.6\linewidth]{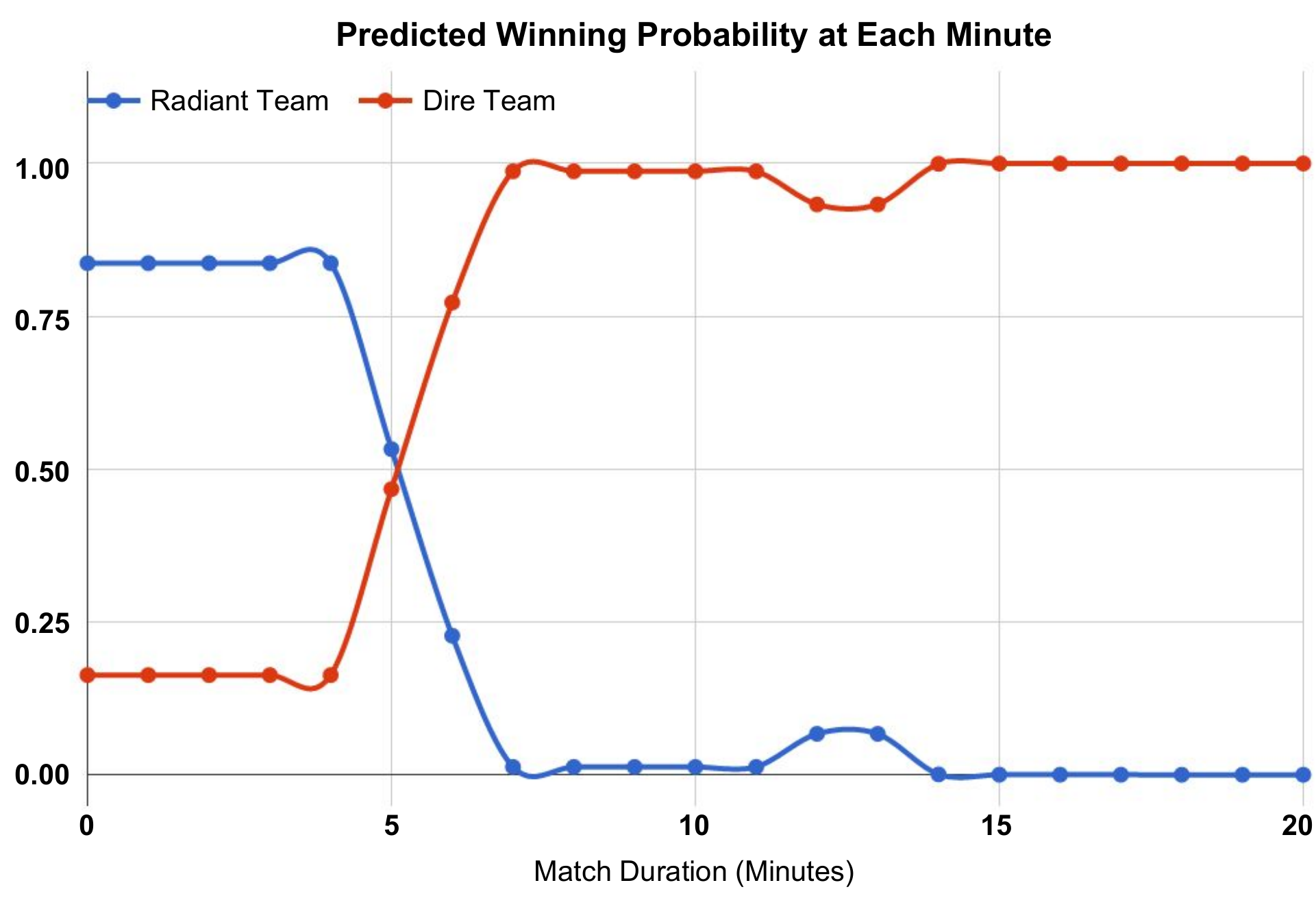}
\label{fig:turnaround_match}
\caption{The winning probability of Radiant team in an example Dota 2 match at each minute predicted by our system. In the beginning of the match, prior information (history data) predicts that the Radiant team has an advantage. However, starting from the 5th minute, real-time information (gameplay data) gains more influence and predicts that the Dire team will reverse the match. In this example, our system is able to achieve nearly 100\% confidence at as early as the 7th minute, long before the match ends.}
\end{SCfigure}

In this paper, the information before a match begins is referred to as \textit{prior} information, and the information during a match is referred to as \textit{real-time} information. Figure \ref{fig:turnaround_match} illustrates an example Dota 2 match predicted by our system. The winning probabilities of both teams are estimated at each minute as the match progresses. At the 0th minute when only prior information is available, our system predicts that the Radiant team has an advantage. However, starting from the 5th minute when real-time information gains more influence, it predicts that the Dire team will reverse the match.


\section{Related Work}
\label{chap:related_work}
There have been numerous efforts on predicting the winning team of a match in different kinds of sports. \cite{KDD2016} proposed a novel probabilistic framework using context information to predict tennis and StarCraft II match results. \cite{BayesianNet} used a Bayesian network with subjective variables on football games. \cite{ELO} used the ELO rating system to derive covariates for prediction models. \cite{CF} used collaborative filtering to predict cases without enough history data.

As for eSports matches, \cite{IEEE-GEM2014} used zone changes, distribution of team members, and time series clustering to investigate spatio-temporal behavior. \cite{UTAH2013} focused on classifying hero positional role and hero identifier based on play style and performance. \cite{FDG2014} worked on discovering patterns in combat tactics of Dota 2. \cite{DERBY2014} trained a neural network to find optimal jungling routes in Dota 2.

Some previous work also tried to predict eSports match results. \cite{ARXIV2015} clustered player behavior and learned the optimal team composition. Then it used team composition-based features to predict the outcome of the game League of Legends. \cite{AASRI2014} predicted the outcome of Dota 2 with topological measures, which are the areas of polygons described by the players, inertia, diameter, distance to the base. \cite{Stanford2013} used logistic regression and k-nearest neighbors to recommend hero selections that would maximize team winning probability against the opponent team. \cite{Stanford2015, UCSD2015} used logistic regression or random forests to predict the winning team of Dota 2. However, the features used in previous work are only hero selection and/or hero winning rates, which are from very limited aspects of the available game data. In addition, the information from the real-time gameplay is entirely ignored. Therefore, an important part of this paper is to expand the feature set that will address these weaknesses.

\section{Dataset}
\label{chap:dataset}
For prior information, we use Dota 2 API\footnote{dota2api: wrapper and parser, https://github.com/joshuaduffy/dota2api} to crawl 78362 matches participated by 19790 players with ‘very high’ skill level\footnote{Matches with very high skill level result in fewer random factors and more resemble professional matches.}. Mean and standard deviation of match duration are 37.75 minutes and 10.42 minutes respectively, which are consistent with real Dota 2 scenario. It is worth noting that the match data in previous work \cite{Stanford2015, Stanford2013, UCSD2015} had much shorter duration (mean is ranged between 23 and 30 minutes) and would lead to more biased prediction results.


Each match contains the winning team, players' account IDs and corresponding hero IDs. In addition, we use other 3rd party APIs to collect statistics of heroes and players. Heroes' data can be obtained from HeroStats API\footnote{HeroStats API, http://herostats.io/}. It contains numerous statistics associated with each hero's abilities such as strength, agility, and intelligence. 

Player's ability (i.e. match history) plays a major role to the outcome of prediction. Opendota API\footnote{OpenDota API, http://docs.opendota.com/} provides a player's statistics, which can be treated as prior knowledge of the player's skill level before a match begins. For example, Dota 2 uses Matchmaking Rating (MMR)\footnote{How Dota 2 MMR Works – A Detail Guide, http://www.dotainternational.com/how-dota-2-mmr-works} as the official ranking score, which estimates the skill level of a player. We could also access a player's match history. Through the past match records, we can know the skill level of a player when they choose a particular hero.

For real-time information, we use Opendota API to collect match replay data such as gold, experience and deaths for all players at each minute. Since not all matches have replays available online, only 20631 out of 78362 matches contain gameplay information in our dataset.


\section{Features}
\label{chap:features}

\subsection{Prior Features}
\label{sec:prior_features}
Feature engineering plays a major role in this paper. Domain knowledge is leveraged to extract useful features from the data. Using the collected data, we categorize the extracted features of the prior knowledge into three types: hero, player, and hero-player combined features. For each match, we obtain features from 10 players in the two opponent teams (Radiant and Dire), and combine them into a single feature vector with Radiant team's feature values in the front half.
Sometimes, a player's statistics can not be accessed if they set their account profile private. In this case, we replace missing feature values with corresponding mean values. Furthermore, we filter matches to guarantee that there are at most two missing players in each match.

\textbf{Hero Feature.} Hero feature $\textbf{x}_{h}$ contains three parts: hero selection, hero attributes, and hero winning rate. Hero selection is binary one-hot encoding indicating which heroes are selected for each team in a match. At the time of writing, there are 113 heroes in the Dota 2. We define the hero selection part as a 113 heroes $\times$ 2 teams = 226-dimensional vector. Hero selection is the only feature implemented in two previous works \cite{Stanford2015, Stanford2013}.


Hero attributes are 26 manually chosen statistics associated with each hero's abilities such as strength, agility, and intelligence. We use a 260-dimensional vector to represent the 10 selected heroes' 26 attributes.

Moreover, we calculate 
the radiant winning rate of a hero against another hero in all history matches. So for each match, there will be 5 Radiant heroes $\times$ 5 Dire heroes = 25 rival hero combinations.

Finally, we concatenate the three parts into a 226 + 260 + 25 = 511-dimensional hero feature vector $\textbf{x}_{h}$. Note that all parts of the hero feature are based on global statistics from all players, independent of the current players' match history.

\textbf{Player Feature.} Player feature $\textbf{x}_{p}$ contains the skills and ranking information about the 10 players. We define player feature $\textbf{x}_{p}$ as a 20-dimensional vector to represent the current 10 players by their MMR scores and MMR percentiles\footnote{For a visualization of the MMR distribution and percentiles among more than one million players, see https://www.opendota.com/distributions}.

\begin{figure}[htbp]
\centering
\begin{minipage}{0.48\linewidth}
\includegraphics[width=\textwidth]{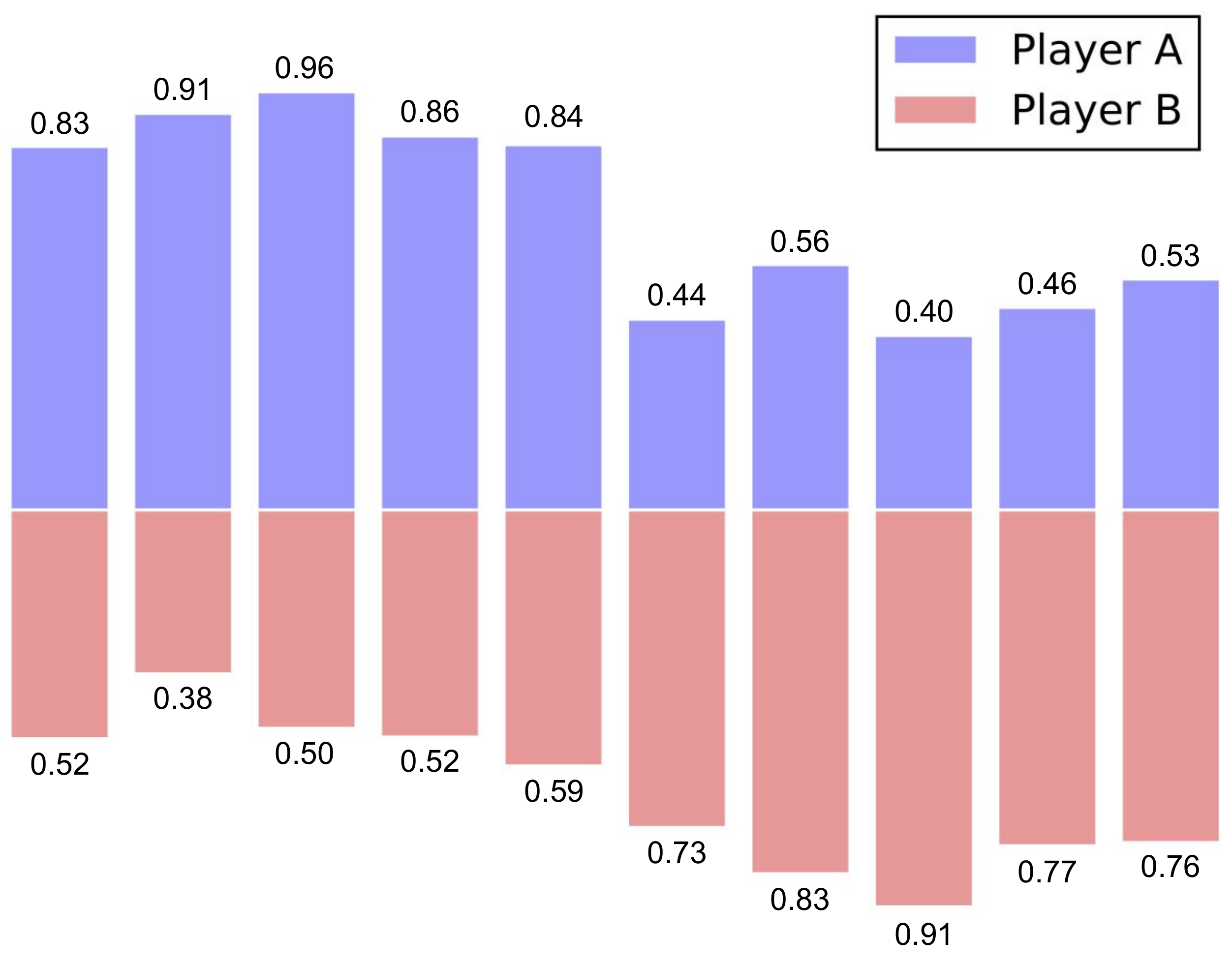}
\caption{Winning rates of two example players when choosing different heroes.}
\label{fig:winrate}
\end{minipage}
\hfill
\begin{minipage}{0.48\linewidth}
\includegraphics[width=\textwidth]{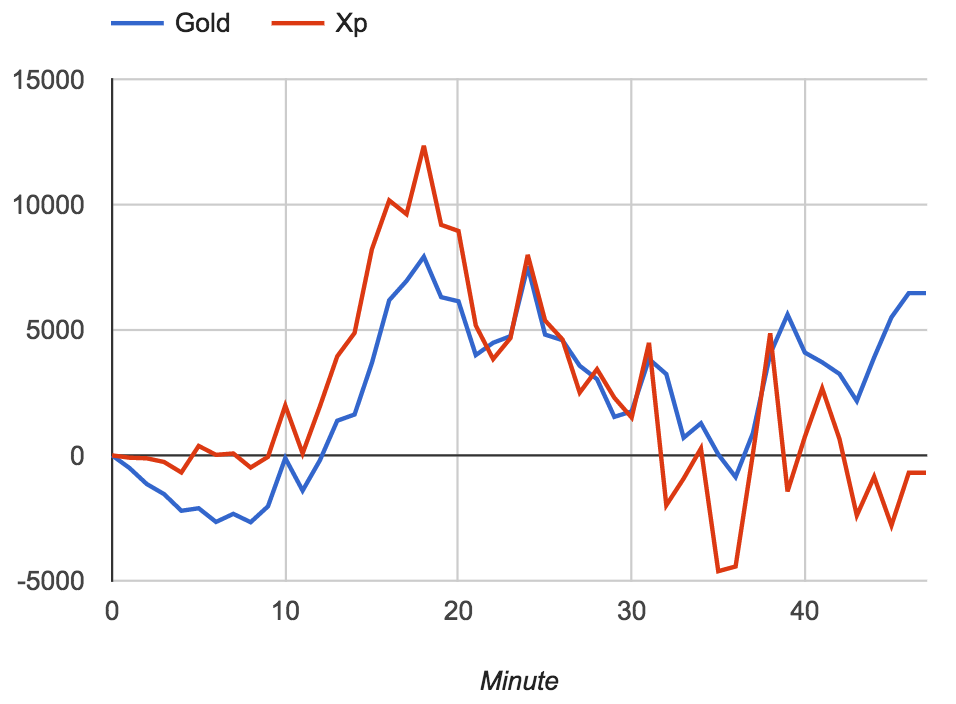}
\caption{Gold and experience difference between the two teams (Radiant minus Dire) at each minute of one particular match.}
\label{fig:time_series}
\end{minipage}
\end{figure}

\textbf{Hero-player Combined Feature.} Hero-player combined feature $\textbf{x}_{hp}$ contains 8 statistics about the 10 players when they choose the current heroes. The statistics include the winning rate when using this hero, mean experience, mean gold gained per minute, and mean number of deaths per minute. We define hero-play combined feature $\textbf{x}_{hp}$ as a 80-dimensional vector to represent 10 players' 8 statistics when choosing the respective heroes for each match. Figure \ref{fig:winrate} shows the winning rates of two example players when choosing different heroes. We can see that player A and player B are good at different heroes with a large variance in winning rates. Such property cannot be captured by hero feature alone since it is averaged over all levels of players. Similarly, player feature alone is not enough since it doesn’t distinguish player’s performance over different heroes.

\subsection{Real-time Features}
\label{sec:realtime_features}
Each team's real-time information can be represented by averaging the five players' gold, experience and deaths at each minute. Then we take the difference between the two teams (Radiant minus Dire) as our feature. Figure \ref{fig:time_series} illustrates the gold and experience features of one particular match. If a match has $T$ minutes, then the real-time feature for this match will have $3 \times T$ dimensions.

\section{Models}
The goal of this paper is to predict the winning team before a match begins (using prior information) and during a match (using real-time information). We refer to these two parts as prior modeling and real-time modeling. In this section, we first introduce how we separately model the prior part and the real-time part. Then we propose two methods to combine prior modeling and real-time modeling.

\subsection{Prior Modeling}
\label{sec:prior_modeling}
For predicting match results before a match begins, we explore the effectiveness of two classifiers: logistic regression (LR) and neural network. Input to the classifiers are the three categories of prior features explained in Section \ref{sec:prior_features}: hero feature, player feature, and hero-player combined feature.

\subsection{Real-time Modeling}
\label{sec:realtime_modeling}
As game progresses, we are able to observe more data related to the performance of each team. In this section, we proposed two methods to model the three time series data (Section \ref{sec:realtime_features}) obtained during the match. 

One intuitive method is to slice time-series data by sliding window and train a LR. For example, we can slice time-series data by 5-minute windows. The LR is trained using all matches' 5-minute-window time series data (convoluted in time domain). When making a prediction at time $t$, we feed time-series data at time $t-5, t-4, ..., t-1$ as features into the LR.

\begin{figure}[htbp]
\begin{minipage}{0.5\linewidth}
\centering
\includegraphics[width=\textwidth]{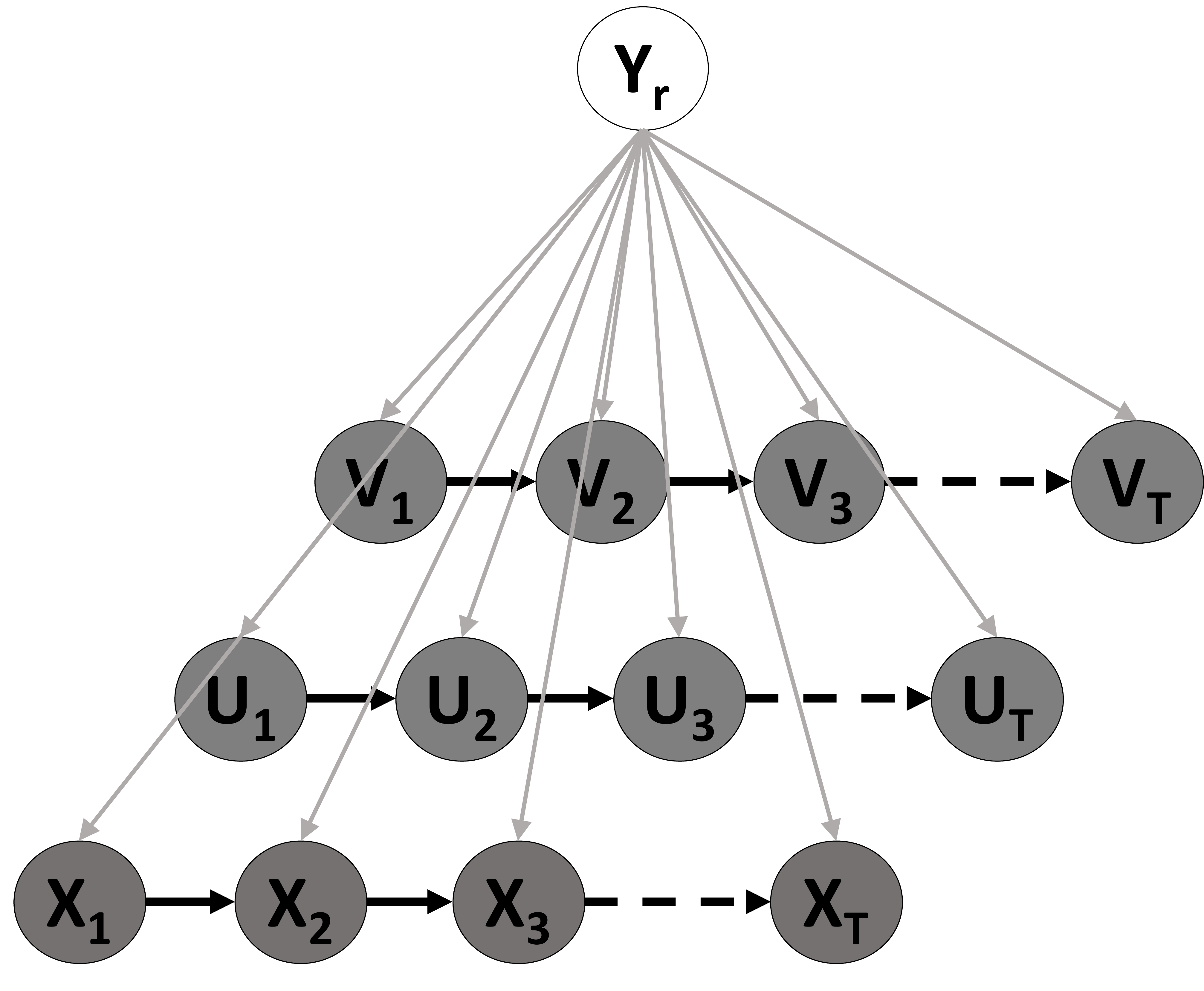}
\vspace{20mm}
\caption{Plate notation of the Attribute Sequence Model (ASM).}
\vspace{-13mm}
\label{fig:model}
\end{minipage}
\hfill
\begin{minipage}{0.35\linewidth}
\begin{minipage}{\linewidth}
\includegraphics[width=\textwidth]{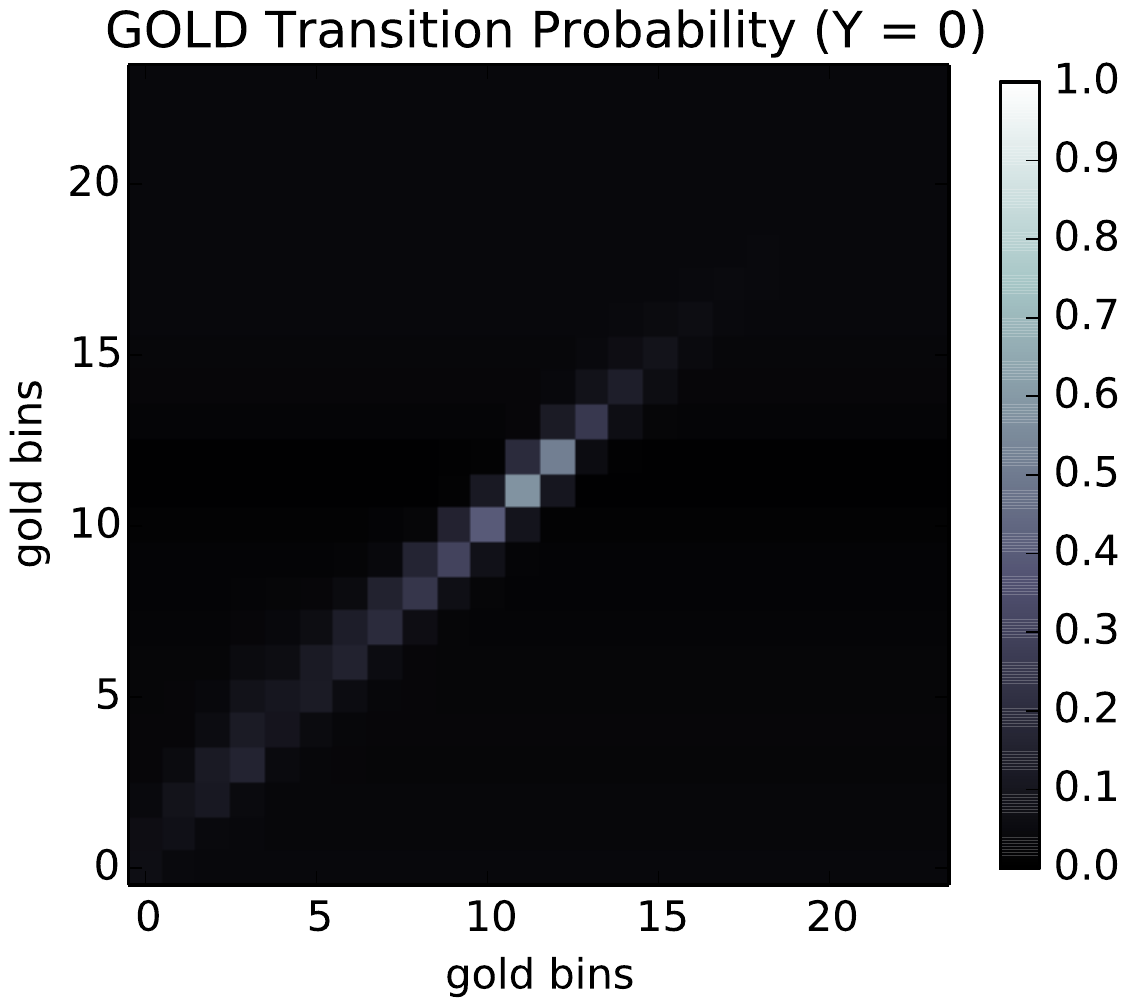}
\label{fig:transition2}
\end{minipage}
\vfill
\begin{minipage}{\linewidth}
\includegraphics[width=\textwidth]{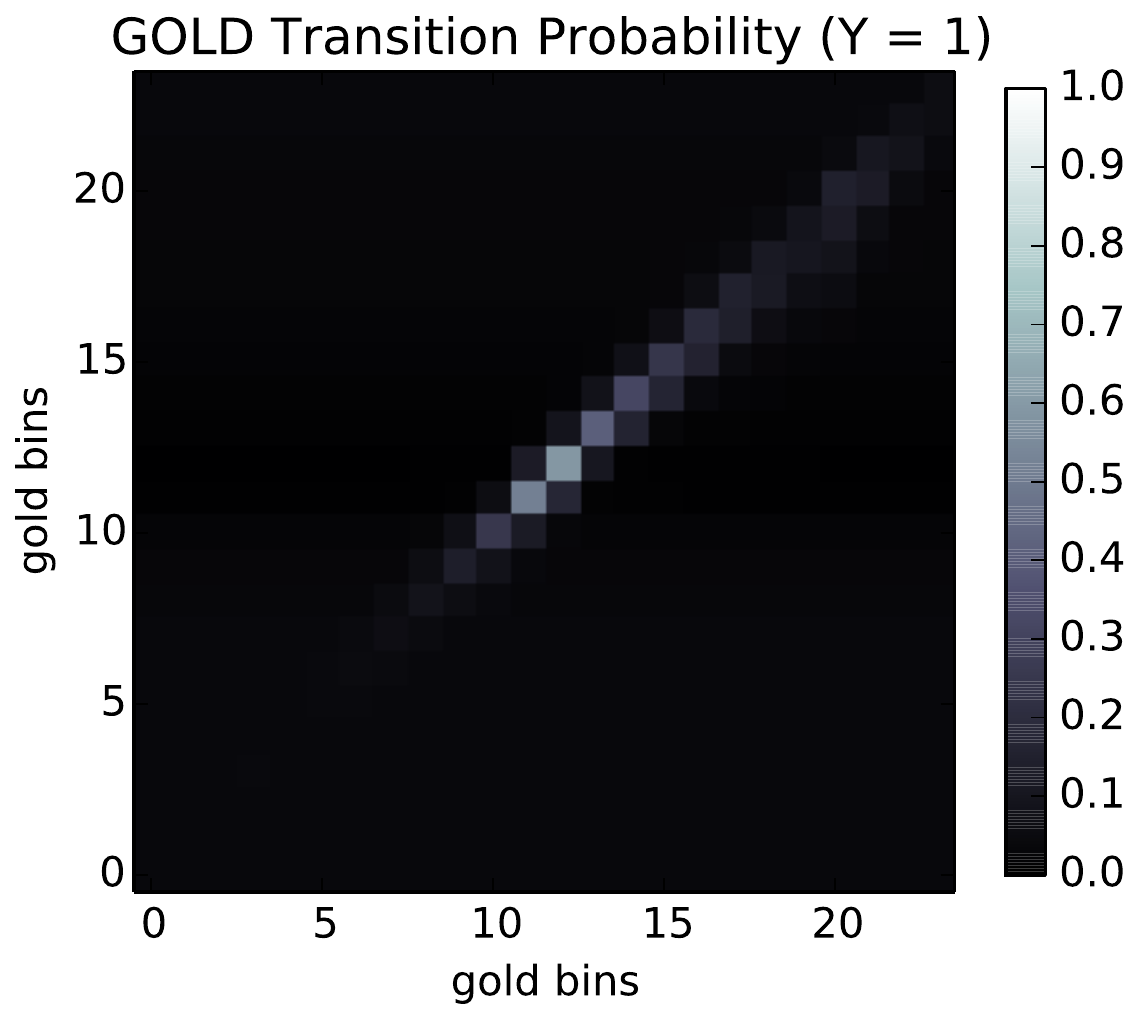}
\label{fig:transition5}
\end{minipage}
\caption{Learned transition probability for the gold time-series data.}
\label{fig:transition}
\end{minipage}
\end{figure}

As the second method, we build a generative model called Attribute Sequence Model (ASM). This model aims to capture the trend of time-series data by explicitly modeling its transition probability in a discrete state space. See Figure \ref{fig:model} for its plate notation.

Here $Y_r$ denotes which team wins the game. $Y_r =1$ if Radiant wins and $Y_r = 0$ otherwise. $X, U, V$ are the three time-series data, with $X_t, U_t, V_t$ represents respectively the deaths, gold and experience at time $t$. Note that here we construct the value of $X_t, U_t, V_t$ as the difference of two teams. For example if at time $t$ Radiant team has mean experience 2000 and Dire team has mean experience 1000, then $V_t = 2000 - 1000$. Further, we discretize $X, U, V$ into bins and assign a discrete bin number to $X_t, U_t, V_t$ rather than the real value. For example, experience 1000 falls into the $8^{th}$ bin thus we set $V_t = 8$. In practice, we divide deaths, gold and experience into 24 bins. Therefore $X_t, U_t, V_t$ can have integer values 0-23.

The generative process of ASM is as follows: First $Y_r$ is sampled to determine which team is going to win. Then at each time $t$, $X_t, U_t, V_t$ are sampled independently based on their previous value $X_{t-1}, U_{t-1}, V_{t-1}$ and $Y_r$. The training process of ASM involves learning transition probability $P(X_t | X_{t-1}, Y_r)$, $P(U_t | U_{t-1}, Y_r)$ and $P(V_t | V_{t-1}, Y_r)$, which are estimated by Maximum Likelihood Estimation. 

Figure \ref{fig:transition} is the visualization of the learned transition probability for gold time-series data. X-axis and Y-axis are bin numbers. The larger the bin number is, the larger the value it represents. Light color represents larger probability. Transition probability can be interpreted as the trend of time-series data under different conditions. For instance, when $Y=1$ which means Radiant is going to win the game, $U$ (Radiant's gold minus Dire's gold) is more likely to transit from a smaller value to a larger value. When $Y = 0$, it is the opposite.

The prediction process of the ASM model is as follows: if we want to make prediction at time $t$, we use the previous 5 minutes' data $D = \{X_{t-5}, ..., X_{t-1}$, $U_{t-5}, ..., U_{t-1}$, $V_{t-5}, ..., V_{t-1}\}$ and calculate the posterior probability  $P(Y_r=y\ |\ D)$, on which our prediction is based.

\subsection{Prior and Real-time Combined Modeling}
\label{sec:prior_realtime_combined_modeling}
To build an accurate prediction model, we need to take into account both prior information and real-time information. In this section we propose two methods to combined prior modeling and real-time modeling together. 

In method 1, we concatenate the 5-minutes-window time-series features with prior features and train a LR. In method 2 we train a new LR on top of the output of prior LR ($Y_p$) and the output of the real-time ASM model ($Y_r$).

\section{Experiments}
\label{chap:experiments}
We split matches into training and testing dataset with ratio of 9:1.  All feature values are normalized to zero mean and unit variance. When obtaining statistics from match history, we exclude matches in the test dataset. Hyper-parameters for LR and neural network are searched via 10-fold cross-validation on training data set. Table \ref{table:cross-validation} is the cross-validation result on selecting hidden size and activation for two-layer neural network. In the end, we using LR with L2 regularization of 1e-6 and two-layer neural network with hidden size of 64 and sigmoid activation. Both classifiers were implemented with Keras\footnote{Keras: Deep Learning library for TensorFlow and Theano, https://github.com/fchollet/keras} library. 

\begin{table}[htbp]
\centering
\caption{Cross-validation accuracy on two-layer neural network for hidden size and activation}
\begin{tabular}{|c|c|c|c|} 
\hline  & 32 & 64 & 128  \\
\hline Sigmoid & 70.48\% & \textbf{70.79\%} & 70.68\%  \\
\hline Relu & 69.06\% & 69.08\% & 68.65\%  \\
\hline Tanh & 69.39\% & 70.04\% & 69.39\%  \\
\hline
\end{tabular}
\label{table:cross-validation}
\end{table}


\subsection{Prediction Using Only Prior Information}
\label{sec:exp_prior}
We compare prediction accuracy with different prior feature combinations by various methods, including two previous works, our logistic regression and neural network. \cite{Stanford2013} trained logistic regression based on only hero selection. \cite{UCSD2015} further added hero against hero winning rate and trained with random forest. We train their features using LR on our dataset.

As can be seen from Table \ref{table:features}, using all features (Hero + Player + Hero-player) achieves the highest prediction accuracy $71.49\%$, which outperforms the two previous works by more than 10\% because they used only subsets of our prior features. Although both previous works claimed to have achieved over 70\% prediction accuracy, their data had much shorter average match duration (between 23 to 30 minutes) that does not reflect the real distribution in the Dota 2 game. Section \ref{sec:discuss_duration} further discusses the effect of match duration on prior prediction.

Furthermore, hero-player combined feature is the single most informative feature, achieving $69.90\%$ prediction accuracy by itself. For example, it includes individual player's performance over different heroes (Figure \ref{fig:winrate}), which is proven to be a key factor in winning a match.

\begin{table}[htbp]
\centering
\caption{Prediction accuracy with different prior feature combinations by various methods.}
\begin{tabular}{|c|c|c|c|c|} 
\hline & \textbf{Hero} & \textbf{Player} & \textbf{Hero-Player} & \textbf{Hero + Player + Hero-Player} \\
\hline Conley et al. \cite{Stanford2013} & 58.79\% & N/A & N/A & N/A \\
\hline Kinkade et al. \cite{UCSD2015} & 58.69\% & N/A & N/A & N/A \\
\hline Logistic Regression & 60.07\% & 55.77\% & 69.90\% & 71.49\% \\
\hline Neural Network & 59.53\% & 56.39\% & 69.71\% & 70.46\% \\
\hline
\end{tabular}
\label{table:features}
\end{table}

\subsection{Prediction Using Both Prior and Real-time Information}
\begin{figure}[htbp]
\centering
\includegraphics[width=0.7\textwidth]{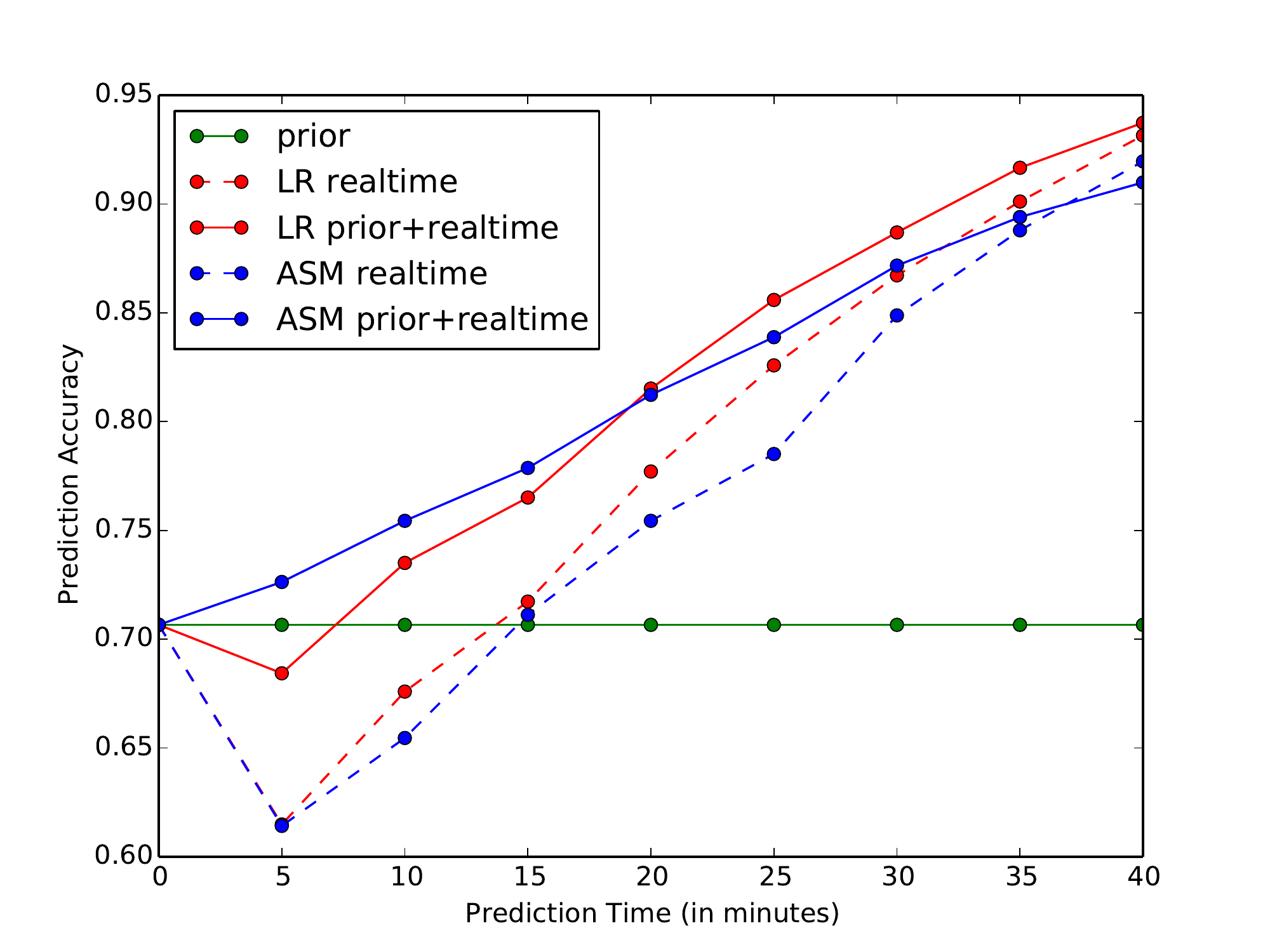}
\caption{Prediction accuracy plot using different models and features. The horizontal axis is the specific minute when prediction is made. Real-time predictions use previous 5 minutes of data.}
\label{fig:acc}
\end{figure}

In Figure \ref{fig:acc}, we compare all the proposed methods in terms of their accuracy when we make prediction at every 5 minutes. The green horizontal line is training LR using only prior features, which serves the baseline of our real-time modeling. The red dashed line is training LR using only real-time features and red solid line is training LR using both prior and real-time features (Section \ref{sec:prior_realtime_combined_modeling} method 1). The blue dashed line is the real-time prediction made by ASM model, and the blue solid line is the combination of ASM model and LR (Section \ref{sec:prior_realtime_combined_modeling} method 2). Three observations can be drawn from these experiments.

\label{sec:exp_realtime}

 \begin{enumerate}
\item Sufficient real-time information greatly improves prediction accuracy over baseline (green line). In our experiments, all four real-time models (blue lines and red lines) achieve better accuracy when fed with real-time information longer than 15 minutes. When predicting at the 40th minutes, accuracy achieves as high as $93.73\%$, in spark contrast to the 69\% accuracy when we only use prior feature to predict at the 40th minute (Figure \ref{fig:duration}).

\item Real-time features become more and more informative than prior features when make prediction at later stage of a match. Notice that the gap between blue solid line and blue dashed line is diminishing. The same happens to red solid line and red dashed line. This suggests that at later stage of a match teams' real-time performance determine the winning side. Prior features lose prediction power as match lasts longer.
\item The ASM approach (blue solid) performs better than LR (red solid) when using real-time features less than 20 minutes. This is probably because at the early stage of a match each team's performance is similar. At this stage it's more important to model the trend of performance rather its current value. ASM model explicitly models the transition probability of the three time-series data which encodes the trend of each team's performance, therefore it outperforms LR at this period.
\end{enumerate}

\section{Discussion and Analysis}
\label{chap:discussion}
\subsection{Effect of Match Duration on Prior Prediction}
\label{sec:discuss_duration}
\begin{figure}[htbp]
\centering
\begin{minipage}{0.48\linewidth}
\includegraphics[width=\textwidth]{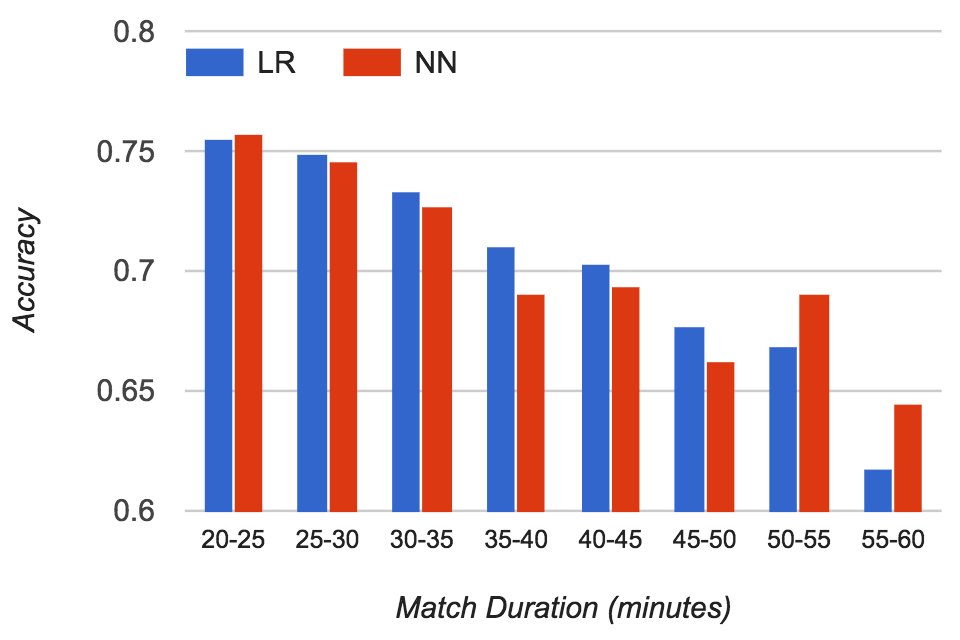}
\caption{Prediction accuracy using only prior information for different match duration.}
\label{fig:duration}
\end{minipage}
\hfill
\begin{minipage}{0.48\linewidth}
\includegraphics[width=\textwidth]{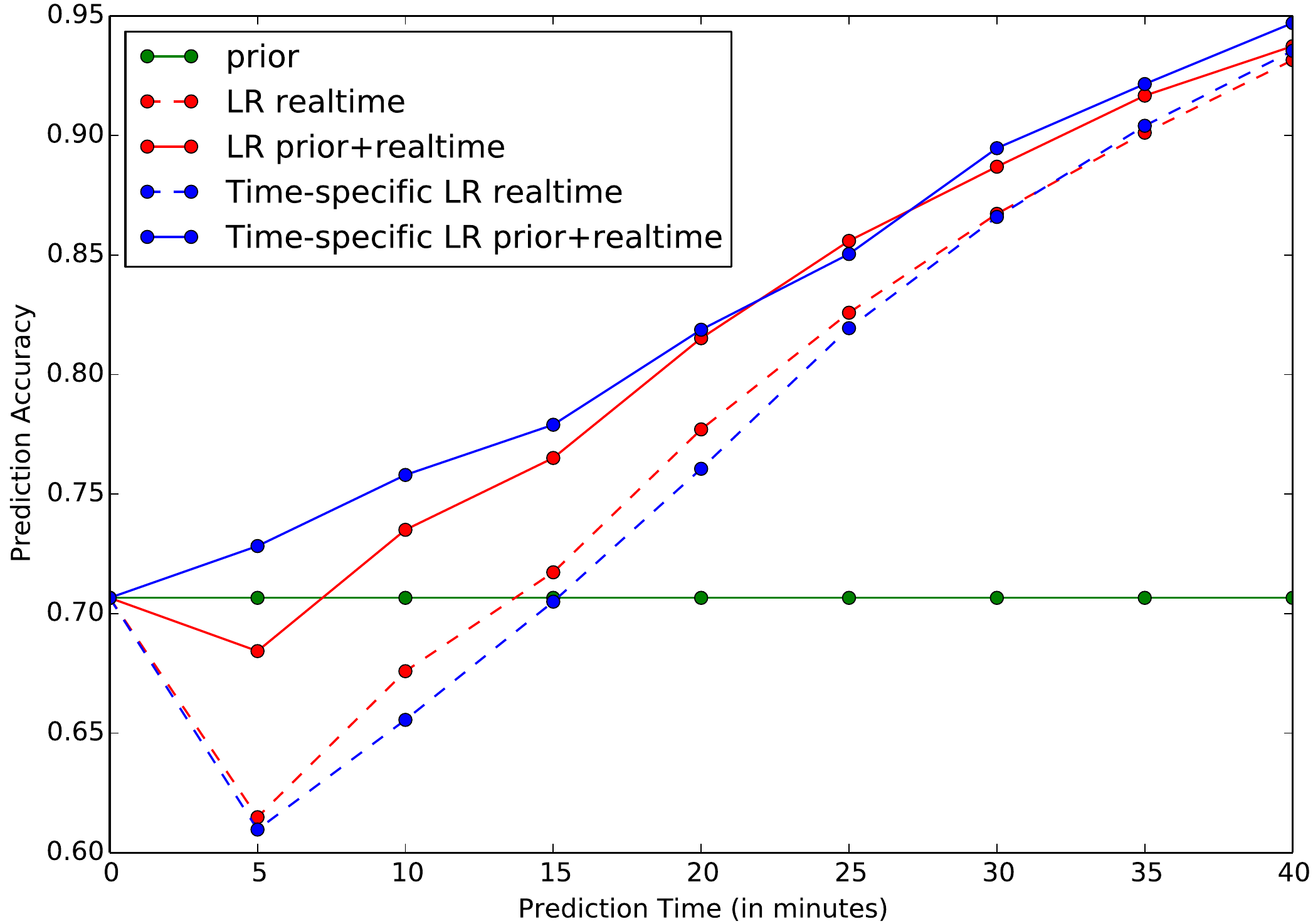}
\caption{Prediction accuracy plot for LR and Time-specific LR.}
\label{fig:multipleLR}
\end{minipage}
\end{figure}

We conduct error analysis to examine the possible cause of wrong predictions when using only prior information (Section \ref{sec:exp_prior}). Figure \ref{fig:duration} shows the effect of different match duration on prediction accuracy. In general, as matches last longer, accuracy drops and the matches become more unpredictable. For matches longer than 55 minutes, only less than $65\%$ accuracy is achieved. This phenomenon is likely due to the fact that as the match progresses, prior information (such as hero selection) is of less importance in its contribution to the final match result. Instead, it is the real-time gameplay information (such as how quickly each hero gains experience) that will give more clues to the final match result. In fact, this hypothesis is also a motivation to introduce real-time gameplay features in our feature set.

\subsection{Time-specific Logistic Regression}
\label{sec:time_specific_lr}
In previous section, we train one LR with 5-minutes-window time-series features. In a match, the model will use multiple 5-minutes-window time-series features starting at different minute. Merging different windows together will lose time-specific information. For example, gaining 1000 gold from 35 minute to 40 minute have different effect comparing to that from 5 minute to 10 minute. Therefore, one way to improve the previous LR is that we train multiple LR. Each LR is trained by time-series features from the same 5-minutes-window. Figure \ref{fig:multipleLR} shows that this time-specific model (Blue solid line) outperforms previous LR (Red solid line) in most time.

\subsection{Assumptions of ASM Model}
\label{sec:assumption_asm}
Here we discuss certain assumptions made in building the ASM model that may be untrue. The ASM model assumes independence of the three time-series. However, we observe strong correlation between them. For example, gold and deaths are highly negatively correlated because in the settings of Dota 2, deaths lead to lost of gold. Also we assume bigram-like transition. These two assumptions are made primarily to reduce model parameters and make it more generalizable. If there are more training data, these two assumptions can be relaxed.

Another important assumption is that for every match, we assume its winning side $Y_r$ remains unchanged throughout the match. In reality, however, it could be for example that Radiant team is about to win during the first half of the game but in the end Radiant loses because it makes an unforeseeable mistake during the second half. The problem is that in the training data we can only observe the final winning side but not the intermediate winning tendency. One possible workaround is to approximate the intermediate winning tendency by features, but we don't explore it in this paper.

\section{Conclusion and Future Work}
\label{chap:conclusion}
In this paper, we predict the winning team of a match in the multiplayer eSports game Dota 2. To address the weaknesses of previous work, we construct a feature set which covers more aspects including hero, player, and hero-player combination as prior features (before a match begins), and the team differences of gold, experience, and deaths at each minute as real-time features (during a match). We explored the effectiveness of LR, the proposed Attribute Sequence Model and their combinations in such prediction problem. Experiment results show that prior features' prediction power drops as the match lasts longer, which is successfully solved by modeling real-time time-series data. We also show that training time-specific LR further improves prediction accuracy by taking into account time heterogeneity.

Future work includes modeling intermediate winning tendency (Section \ref{sec:assumption_asm}) which is a more accurate indicator of game status than final result. Also, more replay data should be used in the real-time modeling, such as hero locations, equipment and major battle events. Lastly, it would be interesting to develop a system that takes in a public match id and automatically visualize the predicted winning probability in realtime.

\bibliography{references}

\end{document}